\documentclass{ws-procs9x6}

\begin{document}

\title{Microscopic foundations of nuclear supersymmetry}

\author{HENDRIK B. GEYER$^1$ and PAVEL CEJNAR$^{1,2}$}

\address{
$^1$Institute of Theoretical Physics, University of Stellenbosch,\\
7602 Matieland, South Africa\\
$^2$Institute of Particle and Nuclear Physics, Charles University,\\
V Hole\v sovi\v ck\'ach 2, 180\,00 Prague, Czech Republic\\
}

\maketitle

\abstracts{
We discuss a microscopic framework for phenomenological boson-fermion
models of nuclear structure based on the U($n/m$) type of superalgebras. 
The generalized Dyson mapping of fermion collective superalgebras 
provides a basis to do so and to understand how collectivity selects 
the required preservation of boson plus fermion number as a good quantum 
number. We also consider the difference between dynamical and invariant 
supersymmetries based on possible supermultiplets of spectra of 
neighboring odd and even nuclei. We point out that different criteria 
exist for choosing the appropriate single particle transfer operators 
in the two cases and discuss a microscopically based method to 
construct these operators in the case of dynamical supersymmetry.
}

\section{Dynamical vs invariant SUSY in nuclear structure}
\label{secintro}

Supersymmetry (SUSY) was originally introduced into relativistic
quantum field theory to exhibit and study invariance with respect to
the exchange of bosons and fermions (see e.g.\ Weinberg's recent
monograph\cite{weinb00} for an overview), but the notion has since
been successfully exploited in a variety of quantum mechanical and
quantum many-body systems (see e.g.\ the review by Cooper 
{\it et al\/}\cite{cooper} and Junker's book\cite{junker96} 
for various examples).

Unfortunately discussions of dynamical supersymmetry on the
phenomenological nuclear structure level, and its relation to the
notion of supersymmetry above, have not always clarified the 
distinction between them, nor that between the concepts of 
dynamical supersymmetry and what we will refer to as \lq invariant 
supersymmetry\rq\ in the nuclear context.  It is therefore not
totally surprising to find that somewhat negative opinions 
such as the following one by 't Hooft\cite{hooft} have been voiced: 
\lq\lq At first sight, the fact that supersymmetric patterns were 
discovered in nuclear physics has little to do with the question 
of supersymmetry among elementary particles, but it may indicate 
that, as the spectrum of particles is getting more and more complex, 
some supersymmetric patterns might easily arise, even if there is 
no `fundamental' reason for their existence.\rq\rq

What we argue below (see also Ref.\cite{cejgey} for a
more complete presentation) is that the situation is much more
positive than this and that the \lq fundamental reasons\rq\ might 
be found (i) in the nature of interactions on the nucleon
level which gives rise to collectivity of nuclear states, and 
(ii) in the utility of boson-fermion mappings which show that 
dynamical supersymmetry can arise in a fermion system without 
violation of the Pauli principle.

{\em Dynamical supersymmetry} on the phenomenological level concerns
situations where states of a quantum system with even and odd fermion 
numbers can be unified in a single representation of a certain 
supergroup. In the nuclear physics context this possibility was 
realized by Iachello\cite{Iac80} in the phenomenological interacting 
boson-fermion model (IBFM)\cite{IvI91} and subsequently shown to be 
applicable to various pairs of nuclei (see Ref.\cite{IvI91} for an 
introduction and review of applications). Renewed interest in this 
possibility has been created by the experimental results of Metz,
Jolie {\it et al\/}\cite{metz99} where a {\it quartet\/} of nuclei 
has been found to fit into a single extended supersymmetric multiplet 
of ${\rm U_{\nu}(6/12)\otimes U_{\pi}(6/4)}$ which takes both 
neutron ($\nu$) and proton ($\pi$) degrees of freedom explicitly 
into account.

Although there is no fundamental difference between even and odd 
nuclei (states of both of them are in principle eigenstates of 
the same Hamiltonian with different particle numbers), their actual
properties differ substantially. Thus unification of spectra 
of even and odd nuclei into a single framework is indeed a challenging 
possibility, with the prospect of unveiling a basic underlying symmetry. 
The notion of supersymmetry has in this regard proved to be quite 
fruitful. It is applied on the phenomenological 
IBFM level, where boson degrees of freedom ($s$- and $d$-bosons) 
are introduced to describe collective monopole and quadrupole 
fermion pairs, while fermions represent only the single (odd) nucleon.   
Dynamical supersymmetry then arises when different boson-fermion 
interaction strengths are related in a special way.

{\it Invariant supersymmetry} in nuclear structure, as considered, 
e.g., by Jolos and von Brentano,\cite{jolos} is much closer in 
spirit and detail to SUSY quantum mechanics.
As is well known,\cite{junker96} the simplest form of invariant 
supersymmetry in quantum mechanics can be formulated in terms of 
nilpotent operators (\lq supercharges\rq ) of the form 
$Q=B\alpha^\dagger$ and $Q^\dagger$ with the $B$'s and $\alpha$'s the 
usual boson and fermion operators, which are defined to be 
kinematically independent: $[B,\alpha]=[B,\alpha^\dagger]=0$.  
The supersymmetric Hamiltonian $\{Q,Q^\dagger\}=B^\dagger B 
+ \alpha^\dagger\alpha$ obviously has eigenstates 
$|n_B,\; n_{\alpha}\rangle$ and displays the hallmark supersymmetric 
spectrum of a unique ground state $|0,0\rangle$ and a set of doubly 
degenerate excited states (this can be easily extended to models 
with more than one supercharge).
An (approximate) invariant supersymmetry in nuclear physics, if 
verified by experimental data, would therefore imply not only the 
above-discussed unique classification of states in even and odd nuclei, 
but also the actual (approximate) degeneracy of some of these states.
As present nuclear data indicate no such a degeneracy, an invariant 
supersymmetry between single nucleons and collective pairs seems to be 
broken into a particular dynamical supersymmetry, quite similarly to
the familiar scenario considered in elementary particle physics.

\section{Dynamical SUSY: phenomenology vs microscopy} 
\label{sec1}

In applications of dynamical supersymmetry to nuclear 
spectra,\cite{Iac80,IvI91,metz99} the appropriate Hamiltonian is
expressed in terms of the above odd generators $Q$ and 
$Q^{\dagger}$, as well as even generators of the type 
$\alpha^{\dagger}\alpha$ and $B^{\dagger}B$.
This selects the U$(n/m)$-type of superalgebras to form
an appropriate algebraic framework describing dynamics of 
the system. In contrast to the invariant supersymmetry, the 
nuclear interaction requires more general terms than those 
appearing in the typical SUSY Hamiltonian above. While the 
invariant supersymmetry is therefore broken, it nevertheless 
remains possible to classify states of some even and odd 
neighbouring nuclei in terms of representations of 
a supermultiplet. This is still a non-trivial property of
the interactions, which must allow for expressing the nuclear
Hamiltonian only in terms of Casimir invariants corresponding 
to a certain chain of (super)algebras that decompose the 
dynamical superalgebra U$(n/m)$.

It is clear that for the whole analysis to be feasible, 
a phenomenological IBFM Hamiltonian which reflects interactions
among bosons and fermions is a necessary prerequisite.
A crucial question thus appears whether the dynamical 
supersymmetry can be compatible with the Pauli principle on 
the microscopic level or, equivalently, whether dynamical 
supersymmetry can be an exact property of a fermion system.  
From the point of view that there are important Pauli corrections 
to the lowest order association between collective fermion pairs 
and IBM bosons,\cite{km91} one might anticipate a negative answer 
to this question.  Nevertheless, the implementation of appropriate 
boson-fermion mappings indeed reveals instances where this
compatibility holds exactly---see also 
Refs.\cite{doba1,navr1,navr2}.

Apart from providing a concrete link between fermion dynamics and
dynamical supersymmetry, the use of boson-fermion mappings also allows
one to {\it construct\/} various transition operators appropriate to the
boson-fermion description, including the important single-particle
transfer operators.  This is in contrast to the phenomenological
situation where one is obliged to truncate an infinite series of
combinations of boson and fermion operators with phenomenological
parameters and terms only restricted by their tensor and particle
number changing properties.\cite{IvI91} It should be emphasized that
the choice of these transition operators in phenomenological models
such as the IBM or IBFM is {\it not} dictated by the Hamiltonian
parameters in general, specifically also in the case of dynamical
symmetry or supersymmetry. (See also Ref.\cite{GND94} for a
discussion of this point.)

\section{Example: dynamical SUSY in the seniority model}
\label{sec2}

The SU(2) seniority model has been analysed exhaustively. Here
we briefly discuss how the known results may be obtained from a
boson-fermion mapping and interpreted from the point of view of
dynamical supersymmetry.

The SU(2) model is defined by considering in a single-$j$
shell the monopole pair creation operator $S^{\dagger}=
\sqrt{\Omega/2}\,(a^{\dagger}_j a^{\dagger}_j)^{(0)}$ with 
$\Omega=j+1/2$.
It fulfills the commutation relation $[S,S^{\dagger}]= 
\Omega-n$, where $n$ is the fermion number operator. The SU(2) 
algebra can be generalized to describe odd systems by constructing
the superalgebra  generated by the operators $S^{\dagger}$, 
$S$, $\Omega$$-$$n$, $a_{jm}$, and $a_{jm}^{\dagger}$. The relevant
commutation relation is $[a_{jm}^{\dagger},S]= 
-\tilde{a}_{jm}$, with $\tilde{a}_{jm}=(-1)^{j-m} a_{j,-m}$, while 
the single-fermion operators obey the standard anticommutation 
relations.  Clearly this 
is a rather trivial superalgebra as the elements of the odd sector 
(single fermion operators) anticommute only to the identity. 
Alternatively, by considering the {\it commutator\/} of single-fermion 
operators, the set of bi- and single-fermion operators may of course 
also be viewed as generators of a standard (orthogonal) algebra.

In the single-$j$ shell we consider the pairing Hamiltonian 
$H=-GS^{\dagger}S$ which has the energy spectrum $E(n,v)=
-\textstyle{\frac{1}{4}}G (n-v)(2\Omega-n-v+2)$, with $n$ the 
total number of fermions and the seniority quantum number
$v$ denoting the number of fermions not coupled to angular 
momentum zero. This Hamiltonian describes both the even and odd 
systems, and the spectra in both cases are given by the same 
expression with $v$ even or odd, respectively.

We can apply to this model the general Dyson boson-fermion mapping
derived in Ref.\cite{navr1} to find an equivalent description in the
boson-fermion space.  As explained in the work cited, the construction, 
which utilises supercoherent states, is a two-step process that 
requires application of a certain similarity transformation, the 
general form of which is derived in Ref.\cite{cejgey}. This procedure 
finally yields typical non-Hermitian Dyson structures that generalize 
the images obtained in the case when only the even sector of the 
collective superalgebra is considered.
Ideal fermion operators $\alpha^\dagger_{jm}$ and $\alpha_{jm}$ 
are now introduced. They obey the standard fermion algebra and
commute with the boson operators $B^\dagger$ and $B$.
Ideal fermion pair operators ${\Sigma}^{\dagger}$ and ${\Sigma}$ 
are obtained from $S^{\dagger}$ and $S$ by replacing all $a$'s 
by $\alpha$'s.  

The mapping obtained for the SU(2) case is
\begin{eqnarray}
S^{\dagger} &\longleftrightarrow&
 B^{\dagger}(\Omega-\aleph)                 , \label{su141c} \\
S  &\longleftrightarrow&
                       B                    , \label{su141a} \\
n &\longleftrightarrow&
 2 N_{\rm B} + N_{\rm F} =
 \aleph + N_{\rm B}                         , \label{su141b} \\
a^{\dagger}_{jm} &\longleftrightarrow&
 \alpha^{\dagger}_{jm}\frac{\Omega-\aleph}{\Omega-N_{\rm F}}
 +B^\dagger{\tilde\alpha}_{jm}-{\Sigma}^{\dagger}{\tilde\alpha}_{jm}
 \frac{\Omega-\aleph}{(\Omega-N_{\rm F})(\Omega-N_{\rm F}+1)}
                                             , \label{su141e} \\
a_{jm} &\longleftrightarrow&
 \alpha_{jm} + 
 {\tilde\alpha}^{\dagger}_{jm} B \frac{1}{\Omega-N_{\rm F}}
 + {\Sigma}^{\dagger} \alpha_{jm} B \frac{1}{(\Omega-N_{\rm F})
 (\Omega-N_{\rm F}+1)}
                                              . \label{su141d}
\end{eqnarray}
Here $N_{\rm F}$ is the number of {\em ideal\/} fermions,
$N_{\rm B}$ the number of bosons, and $\aleph= N_{\rm F}+N_{\rm B}$.
Note that the finiteness of the original single-$j$ Hilbert
space implies the necessity to cut off a spurious
sector from the ideal boson-fermion space. In the present case,
the {\em physical subspace\/} satisfies the conditions
$N_{\rm F}\leq\Omega$ and $N_{\rm B}\leq\Omega-N_{\rm F}/2$.
We see that the single fermion images (\ref{su141e}) and
(\ref{su141d}) are finite and contain terms changing the ideal 
fermion number by one only. Furthermore, they preserve
{\em exactly\/} the anticommutation relations on the full ideal 
space, i.e., as operator identities. This property, guaranteed 
by the construction, ensures the exact preservation of the Pauli 
exclusion principle once the original fermion problem is mapped 
into the boson-fermion space.

The mapping above transforms the two-body Hamiltonian 
$H=-GS^{\dagger}S$ 
into a one- plus two-body boson-fermion Hamiltonian of the form
$H_{\rm BF}=-GN_{\rm B}(\Omega-N_{\rm B}+1-N_{\rm F})$.
$H$ and $H_{\rm BF}$ have exactly the same spectrum, 
$E(n,v)=E(N_{\rm B},N_{\rm F})$, which can be seen explicitly by 
equating particle numbers in the two formulations, 
$n=2N_{\rm B}+N_{\rm F}$, and associating $v=N_{\rm F}$.
$H_{\rm BF}$ can also be expressed in a form which stresses its 
dependence on the total number of bosons and fermions, $\aleph$, 
i.e., $H_{\rm BF}=-G(\aleph-N_{\rm F})(\Omega+1-\aleph)$.
Note that the boson-fermion interaction term, $GN_{\rm B}N_{\rm F}$,
can be expressed in terms of the odd generators, $O^{\dagger}_m=
\alpha^{\dagger}_{jm}B$ and $O_m=B^{\dagger}\alpha_{jm}$, of the 
U(1/2$\Omega$) superalgebra. Since the boson and ideal fermion 
number operators can be linked to even generators, it is possible 
to write $H_{\rm BF}$ in yet another form in terms of both
even generators and supergenerators of U$(1/2\Omega)$:
\begin{equation}\label{pairhbfb}
H_{\rm BF}=-G\biggl[N_{\rm B}(\Omega-N_{\rm B}+1)
    +N_{\rm F}-\sum_m O^{\dagger}_m O_m\biggr] \; .
\end{equation}

\section{Preservation of the total number of bosons and fermions}
\label{sec3}

%This clearly illustrates how the degrees of freedom appropriate to the 
%description of a fermion system in terms of supersymmetry, may be 
%systematically introduced. 
%The total number of 
%ideal particles are accordingly preserved as utilized in the 
%classification of states of phenomenological dynamical SUSY. (We 
%return to the general status of this point in the next section.)

Above we have given an example of how the appropriate boson and 
fermion degrees of freedom might enter, without violation of the Pauli 
principle, in a dynamical supersymmetric description of the states of 
a fermion system. However, since the real fermion number maps onto the 
number of ideal fermions plus {\it twice\/} the number of bosons, 
see Eq.\ (\ref{su141b}), it is clear that preservation of the real
fermion number alone cannot explain the appropriateness of the
labelling of dynamical SUSY states by the total number of ideal 
particles in the generalized IBFM description.

On the other hand, in the above example it can be seen explicitly 
that the boson-fermion Hamiltonian (\ref{pairhbfb}) {\em does\/}
indeed preserve the total number $\aleph$ of ideal particles.
Quite recently we have shown, see Ref.\cite{cejgey}, that in the
most general case this property follows from the fact that 
the original fermion Hamiltonian can be expressed in terms of 
operators belonging only to the collective algebra, i.e., operators
$S^{\dagger}$, $S$, and $n$ in the above example. It is of 
course well known that this situation occurs exactly in all the 
well-studied algebraic models and one therefore suspects that 
(approximate) decoupling of a subset of collective states for 
realistic interactions leads to mapped boson-fermion Hamiltonians 
that can indeed be represented by the U$(n/m)$-type of dynamical
superalgebras. 

Moreover, as the above general condition requires the numbers of 
ideal fermions and bosons, $N_{\rm F}$ and $N_{\rm B}$, to be 
conserved {\em separately}, the collectivity in the original 
fermionic problem selects also a class of appropriate dynamical 
symmetries on the mapped boson-fermion level. These symmetries 
must involve the decoupling of the boson and fermion degrees 
of freedom in the first step of the corresponding superalgebraic 
chain, i.e., they must be of the form ${\rm U}(n/m)\supset 
{\rm U}_{\rm B}(n)\otimes{\rm U}_{\rm F}(m) \supset\dots$, as
verified by experimental data.\cite{metz99} This positive 
finding has also a somewhat disappointing aspect, namely that
from the purely spectroscopic viewpoint the nuclear supersymmetry 
seems to be limited just to the possibility to simultaneously 
describe neighboring even and odd nuclei by a boson-fermion 
(IBFM) Hamiltonian with the same set of parameters---a situation 
which cannot be regarded as very surprising.

\section{Single particle transfer operators in nuclear SUSY}
\label{sec4}

It is important to realize that the phenomenological SUSY analysis 
depends non-trivially on the choice of single-particle 
transfer operators and that this choice is not dictated by the 
Hamiltonian which exhibits dynamical 
supersymmetry in a given case. On the other hand, from the microscopic 
viewpoint it is clear that the appropriate transfer operators to be 
used in conjunction with states classified according to 
representations of a dynamical superalgebra are boson-fermion images 
of the original single-fermion operators. This is simply so because 
single-fermion operators are the physical operators associated with 
single nucleon transfer. 

Even the simple SU(2) seniority model discussed above illustrates 
interesting consequences for the structure of transfer operators.
As we see in Eqs.~(\ref{su141e}) and (\ref{su141d}), these 
operators acquire terms which are responsible for the Pauli 
correlations between the even core and the odd particle. While, 
as discussed above, in the phenomenological supersymmetric models 
these terms are postulated or motivated semi-microscopically, the 
SUSY picture derived from the boson-fermion mapping yields transfer 
operators fixed by the mapping procedure itself. The image of the 
fermion annihilation operator (\ref{su141d}) is, e.g., a combination 
of the ideal fermion annihilation operator and two corrective terms.
It should be stressed here that although the apparent non-Hermicity
of the Dyson images seems to obscure their use on the phenomenological
level, it is in fact possible to calculate all the relevant 
single-particle (and collective-pair) transfer matrix elements
with no explicit reference to the microscopic structure of the
collective states involved. This was discussed by the present
authors in detail in Ref.\cite{cejgey}.
When the microscopic nucleon level interaction is more involved than 
the schematic pairing case studied above, the images should of course 
be generalized along the lines of the seniority mapping as employed by 
Navr{\'a}til and Dobe{\v s},\cite{navdob1,navdob2} and by Geyer and 
Morrison.\cite{geymor} 

On this issue we therefore disagree with the position of Barea 
{\it et al\/}\cite{barea} who indicate that single particle transfer 
should \lq\lq theoretically (be) described by the supersymmetric 
generators that change a boson into a fermion and {\it vice 
versa},\rq\rq\ 
i.e., by the supercharges $Q$ and $Q^{\dagger}$ considered earlier. 
The supercharges can only be the appropriate transfer operators if, 
for a given set of states, some of the (operator dependent) 
coefficients in expressions such as (\ref{su141e}) and (\ref{su141d}) 
are suppressed to the extent that only the supercharge components are 
effective. This special situation is of course quite interesting in 
itself, and should be further explored within the framework of 
invariant supersymmetry in the sense of \lq fingerprint\rq\ 
degenerate spectra, as discussed, e.g., by Jolos and von 
Brentano.\cite{jolos} The realization of an invariant supersymmetry 
with 
supercharge transfer operators, which is only a hypothetical 
possibility at present, would indeed bring the nuclear SUSY very 
close to the original notion of supersymmetry in elementary
particle physics.

\section{Conclusions}

We have shown how a microscopically based framework for {\it dynamical
supersymmetry} in nuclei arises, taking care to distinguish this
situation from {\it invariant supersymmetry} for which far less 
evidence exists.  In particular, the microscopic origin has been 
discussed of the total number of ideal particles (fermions plus bosons) 
as a good quantum number in dynamical supersymmetry. Amongst other 
things, resolving this issue implies that the boson-fermion mapping 
facilitates the identification of dynamical supersymmetry in 
a fermion system which may be perfectly compatible with all Pauli 
restrictions. We have also discussed the basis for constructing 
single particle transfer operators for the respective manifestations 
of supersymmetry above, pointing to the way ahead for calculations 
of single particle strengths in dynamical supersymmetry.

%\section*{Acknowledgments}
This work was supported by the South African National
Research Foundation under grants GUN 2047181 and GUN 2044653 and
partly by the Grant Agency of Czech Republic under grant
202/02/0939.

\end{document}